\documentclass[twocolumn,10pt]{tsfp}
\usepackage{flushend}
\usepackage{graphicx}
\usepackage[authoryear,round]{natbib}
\usepackage{fancyhdr}
\usepackage{caption}
\usepackage{subcaption}
\usepackage{bm}
\usepackage{epstopdf, epsfig}
\usepackage{amsmath}
\usepackage{amssymb}
\usepackage{mathtools}
\usepackage{bm}
\usepackage{float}
\usepackage{lipsum}
\usepackage{color}

\title{Higher DNS-resolution requirements for expanded overlap region
and confirmation of a convergence criterion}

\author{Sergio Hoyas
          \affiliation{IUMPA\\
	Universitat Politècnica Val\`encia\\
	Camino de Vera, 46024 Val\`encia, Spain\\
	serhocal@mot.upv.es
    }
}

\author{Ricardo Vinuesa%
\affiliation{
 FLOW, Engineering Mechanics,\\
 KTH Royal Institute of Technology,\\
 SE-100 44, Stockholm, Sweden\\
 rvinuesa@mech.kth.se
}%
 }

\author{Peter J. Schmid
    \affiliation{
	Department of Mechanical Engineering\\ King Abdullah University of \\
 Science and Technology (KAUST),\\
    Thuwal, Saudi Arabia\\
    peter.schmid@kaust.edu.sa
    }	
}

\author{Hassan Nagib
    \affiliation{
	MMAE Dept., Armour College of Engineering\\ ILLINOIS TECH (IIT),\\
    Chicago, IL 60616\\
    nagib@iit.edu
    }	
}

\begin{document}

\maketitle   
\thispagestyle{fancy}

\fontsize{9}{11}\selectfont

\section*{ABSTRACT}
Direct numerical simulations (DNS) stand out as formidable tools in studying turbulent flows. Despite the fact that the achievable Reynolds number remains lower than those available through experimental methods, DNS offers a distinct advantage: the complete knowledge of the velocity field, facilitating the evaluation of any desired quantity. This capability should extend to compute derivatives. Among the classic functions requiring derivatives is the indicator function, $\Xi(y^+) = y^+\frac{{\rm d}\overline{U}_x^+}{{\rm d}y^+}$. This function encapsulates the wall-normal derivative of the streamwise velocity, with its value possibly influenced by mesh size and its spatial distribution. The indicator function serves as a fundamental element in unraveling the interplay between inner and outer layers in wall-bounded flows, including the overlap region, and forms a critical underpinning in turbulence modeling. Our investigation reveals a sensitivity of the indicator function on the utilized mesh distributions, prompting inquiries into the conventional mesh-sizing paradigms for DNS applications.
\section*{Introduction}
Since the early work by ~\cite{kim87}, direct numerical simulation (DNS) has become a widely used approach to study fundamental aspects of wall-bounded turbulence~\citep{spalart,hoy06,sillero,wing,pirozzoli,alc21b,hoy22}. Due to its profound implications, the overlap region of the mean velocity profile in wall-bounded turbulence has emerged as a focal point of considerable interest within the research community (See Figure 1). Here, the coexistence of large structures from the outer layer with smaller ones significantly affected by wall and viscous effects are observed \citep{tow76,jim08}. Classical literature asserts that the mean velocity profile in this overlap region adheres to the renowned logarithmic law, characterized by its K\'arm\'an constant ~\citep{vonkarman,millikan}, $\kappa$. This formula reads:

\begin{equation}
    \overline{U}_x^+(y^+) = \frac{1}{\kappa} \log(y^+) + B.
\end{equation}
In this equation, $B$ is the interception constant, $\overline{U}_x$ is the mean streamwise velocity, and $y$ is the distance to the closest wall. All the quantities with $+$ superscripts have been nondimensionalized using the viscosity, $\nu$, and the friction velocity, $u_\tau$. 

\begin{figure}[ht]
\centering
\includegraphics[width=0.50\textwidth]{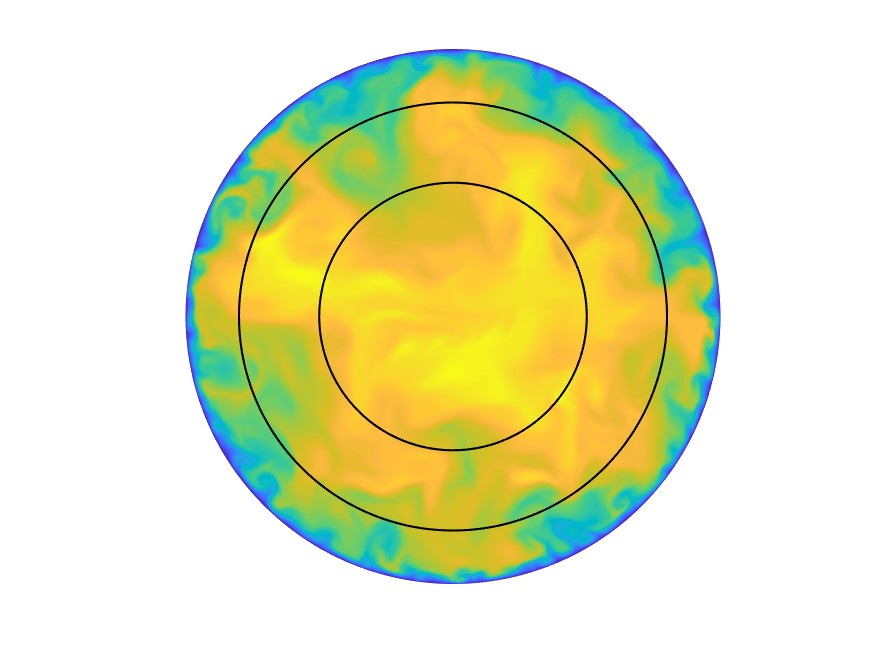}
\caption{$r-\theta$ view of the instantaneous streamwise component $U_x$. The flow has been nondimensionalized by the maximum velocity. The scale goes from approximate $0$ (blue) to 1 (light yellow). The black lines indicate the boundaries of the overlap region, shown here at $y/\delta = 0.2$ and $0.5$.}
\label{fig:figure1}
\end{figure}

\begin{figure}[ht]
    \centering
    \begin{subfigure}{0.49\textwidth} 
    \includegraphics[width=0.99\textwidth]{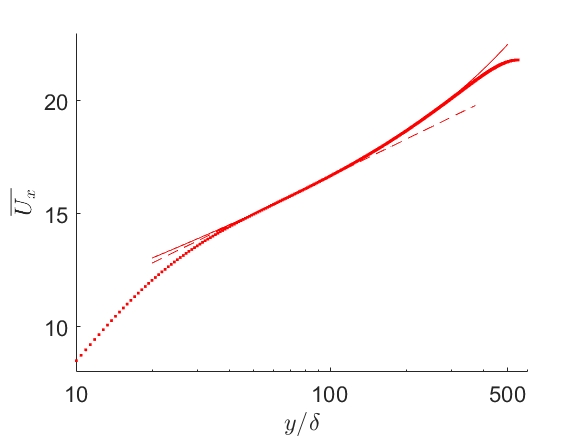}
    \end{subfigure}
    \begin{subfigure}{0.49\textwidth} 
    \includegraphics[width=0.99\textwidth]{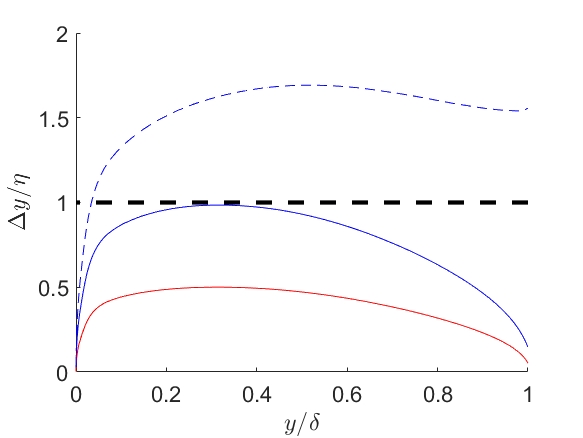}
    \end{subfigure}
        \caption{(Top) Streamwise velocity for the PHRF case (thick line), together with the pure log (dashed line) and log-plus-linear (thin line) overlap layers. (Bottom) Grid distance in the wall-normal direction, scaled with the Kolmogorov length. Colors as in Table~1, where solid lines denote pipe and dashed ones channel. }
    \label{fig:fig1}
\end{figure}

Over the years, this has been a topic of active study, and the universality of the log law and the von K\'arm\'an coefficient has occasionally been challenged or reaffirmed~\citep{george,barenblatt,nagib_debate,vinuesa_exp,variations,luchini,obe22}. Recently,  ~\cite{mon23} (MN from now on) shed additional light on this topic, challenging the accumulated knowledge during the last century. MN's main point is to consider in the inner asymptotic expansion a term proportional to the wall-normal coordinate, $\mathcal{O}(Re_{\tau}^{-1})$. Here $Re_{\tau} = u_\tau \delta / \nu$ is the friction Reynolds number with $\delta$ a characteristic outer length. This $\delta$ can be thought of as the radius $R$ in pipes, the semi-height in channels, $h$, or a length scale related to $\delta_{99}$ in boundary layers. Following MN, we find that the pure log law is not observed in channels and pipes (as well as other flows with streamwise pressure gradients) even if $Re_{\tau}$ is not above $\mathcal{O}(10^4)$. MN's extended matched asymptotics reveals an additional term in the expansion of the velocity for this layer, leading to an extra contribution in the overlap layer in the form $S_0 y^+ Re_{\tau}^{-1}$, where $S_0$ is a coefficient. 
\begin{figure}[ht]
    \centering
    \begin{subfigure}{0.49\textwidth} 
    \includegraphics[width=0.99\textwidth]{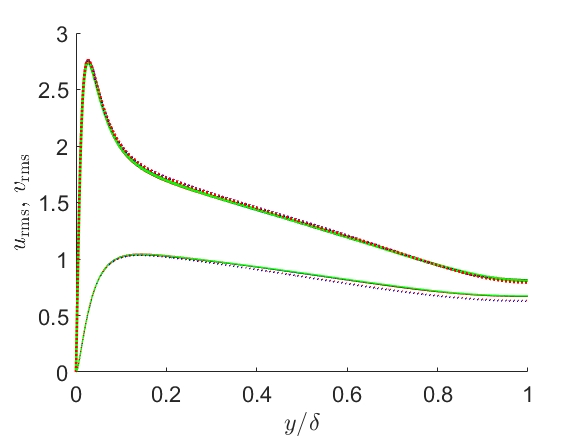}
    \end{subfigure}
    \begin{subfigure}{0.49\textwidth} 
    \includegraphics[width=0.99\textwidth]{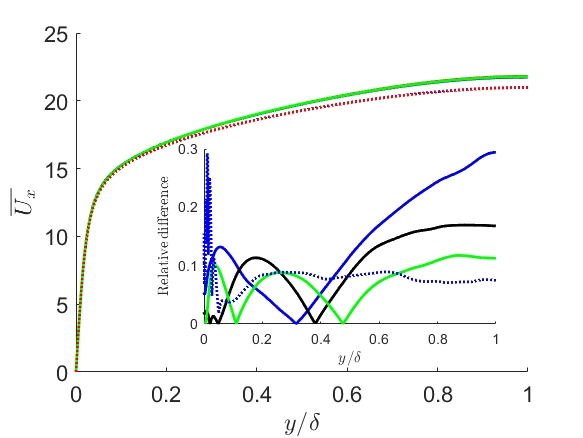}
    \end{subfigure}
    \caption{(Top) Intensities in the wall-normal and streamwise directions for all cases. (Bottom) $\overline{U}_x$. The inset shows the difference between the fine-resolution case and the others. Colors as in Table 1. Continuous lines, pipes and dotted lines, channels}
    \label{fig:fig2}
\end{figure}
Interestingly, the overlap region represented by a combined logarithmic and linear terms exhibits values of the von K\'arm\'an coefficient $\kappa$ consistent with those obtained from skin-friction relations, a fact that suggests that this approach can reveal high-$Re$ trends even at moderate Reynolds numbers. The main point of this matched asymptotic theory is to use a new expression in the overlap region for $\overline{U}_x$. MN's formula reads: 

\begin{equation}\label{eq:010}
\overline{U}_{x,\rm in}^+(y^+\gg 1) \sim \kappa^{-1}\ln y^+ + B_0 + B_1/Re_\tau +S_0  y^+ /Re_\tau.
\end{equation}

\begin{table*}
\centering

\begin{tabular}{lcccccccccccc}
\hline\hline
Case & $ Re_{\tau } $ & $L_{x}/\delta$ & $L_{z}/\delta$ & $\Delta x^{+}$ & $%
\Delta (R\theta ^{+})$ & $\max (\Delta y^{+}$) & $\min (\Delta y^{+})$ & $%
N_{x}$ & $N_{z}$ & $N_{y}$ & ETT & $\varepsilon $ \\ \hline\hline
{\color{blue} PLRF} & 549 & $10\pi$ & $2\pi$ & 5.62 & 3 & 3.2 & 0.018 & 3072 & 1152 & 256 & 
93 & $1.2\times 10^{-4}$  \\ 
{\color{red}  PHRF} & 549 & $10\pi$ & $2\pi$ & 5.62 & 3 & 1.6 & 0.005 & 3072 & 1152 & 512 & 
87 & $2.6\times 10^{-4}$ \\ 
{\color{black} PLRC}& 549 & $10\pi$ & $2\pi$ & 11.2 & 4.5 & 3.2 & 0.018 & 1536 & 768 & 256 & 
184 & $8.2\times 10^{-5}$ \\ 
{\color{green} PHRC}& 549 & $10\pi$ & $2\pi$ & 11.2 & 4.5 & 1.6 & 0.005 & 1536 & 768 & 512 & 
56 & $3.8\times 10^{-4}$ \\ 
{\color{blue} CLRC} & 546  & $8\pi$ & $3\pi$ & 9 & 5 & 5.85 & 0.8 & 1536 & 1152 & 251 & 150 & $7.6\times 10^{-5}$ \\ 
{\color{red} CHRC} & 546 & $8\pi$ & $3\pi$ & 9 & 5 & 1.68  & 0.05 & 1536 & 1152 & 901 & 150 & $3.8\times 10^{-5}$ \\ \hline\hline
\end{tabular}
\caption{Summary of the simulations discussed in this work, where `P' stands for pipe and `C' for channel. The nomenclature `LR' and `HR' stands for low and high resolution in $y$, whereas `C' and ´F' denote coarse and fine in $x$ and $z$. Note that $L_x$ and $L_z$ are the periodic streamwise and spanwise dimensions while $h$ is the channel half height. $\Delta x^{+}$ and $\Delta z^{+}$ are the inner-scaled resolutions in terms of dealiased Fourier modes. The wall-normal direction is indicated in both flows by $y$. $N_x, N_z,$ and $N_y$ are the numbers of collocation points in the three different directions. The time span of the simulation is given in terms of eddy turnovers $u_\tau T/\delta$, and $\varepsilon$ is a measure of convergence, defined in \cite{vin16}. Colors given in the first column are used throughout the paper, while using dashed lines for the channel.}
    \label{tab:literature}
\end{table*}
This model is depicted in Figure 2(top) as a thick line. The classic logarithmic is displayed using a thin line. However, to obtain accurate values of these coefficients, it is necessary to address two fundamental issues:
\begin{enumerate}
    \item  To find $\kappa$, the challenge is to identify the location and extent of the overlap region, which is weakly dependent on the Reynolds number \citep{mon23}.
    \item  To obtain the correct value of $S_0$, it is necessary to obtain with high accuracy the wall-normal distribution of ${\rm d \overline{U}_x^+}/{\rm d}y^+ $.
\end{enumerate}

Focusing on the second point above, it is possible to obtain an equation for the value of $\kappa$ and $S_0$ through the indicator function: 
\begin{equation}\label{eq:020}
 \Xi(y^+) = y^+\frac{{\rm d}\overline{U}_x^+}{{\rm d}y^+}.   
\end{equation}

From equations (\ref{eq:010}) and (\ref{eq:020}), one can obtain the equation for $\kappa$ and $S_0$:
$$
\Xi(y^+) = \kappa^{-1} +S_0  y^+ /Re_\tau.
$$

Thus, to accurately determine $S_0$, it is necessary to compute the indicator function with high accuracy, an aspect that has not been sufficiently investigated in the literature. While a convergence criterion exists for fully developed flows \citep{vin16}, and the domain size of the problem has been examined in depth~\citep{loz14,llu18}, the necessary grid spacing has received much less attention. Examples can be found in some very large simulations \citep{hoy22,yao23}, where the authors only report their mesh size and compare their results with those in previous references, which basically do the same.

We have performed two different numerical experiments to examine this issue and its implications. The details can be found in Table 1. On the one hand, we have studied a smooth pipe of radius $R$ and length $L_x=10\pi R$ for $Re_\tau=550$. We will use $y$ to indicate the distance to the wall. We have used four different meshes, as seen in Table 1. The base mesh is the case PLRC with the same resolution in $x$ and $\theta$ as a channel. On the other hand, the mesh in the radial direction has twice the number of points needed in a channel due to the skewness of the cells near the pipe center. The mesh distribution can be seen in Figure 2 (bottom). Note that for the pipe cases, the wall-normal grid spacing is below the Kolmogorov scale throughout the computational domain. Starting with the PLRC, we doubled the cells in the radial direction (PHRC), azimuthal, and streamwise directions (PLRF), and all of them (PHRF). All these simulations ran for at least 55 eddy-turnover times (ETT), defined as ${\rm ETT}=T u_\tau/R $. This is the primary motivation behind making our initial phase of DNS  to $Re_\tau \sim 550$, as these cases can be run relatively quickly. 

On the other hand, we have also run two channel-flow simulations. The first one (CLRC) is a simulation with the same mesh as Jim\'enez and Hoyas \citep{hoy08,jim08}. The second one (CHRC) uses a mesh considered enough for $Re_\tau=3000$~\citep{loz14,alc21b}. Two different codes were employed. OpenpipeFlow was used to run the pipe simulations \citep{openpipeflow,yao23}. The code LISO was used for running the channel flow \citep{hoy06,llu21c,yu2022three}. Both codes employ Fourier-decomposition techniques in the streamwise and spanwise directions. Openpipeflow uses a seventh-order finite-difference scheme in the wall-normal direction, while LISO uses a tenth-order compact-finite-difference scheme \cite{lel92,llu21c}. 

\begin{figure}[ht]
\centering
\includegraphics[width=0.50\textwidth]{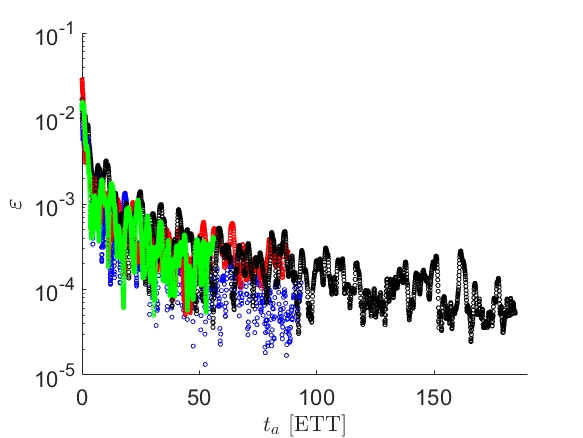}
\caption{Evolution of the error in the momentum equation (\ref{eq:totalshearstress}) integrated over the cross section, as a function of the averaging time $t_a$ expressed in eddy-turnover times. }
\label{fig:epsilon2}
\end{figure}

One of the tools used to assert whether a statistically steady state has been reached is to compute the total shear stress, as displayed in Figure~\ref{fig:epsilon2}. This equation is 
\begin{equation}
1-y^+=\frac{\mathrm{d}\overline{U}_x^+}{\mathrm{d}{y}^+}-\overline{uv}^+.
\label{eq:totalshearstress}
\end{equation}

\noindent where $\overline{uv}^+$ is the Reynolds stress. The parameter $\varepsilon$ is defined as the $L_2$ norm of the error between the left and right-hand sides of equation \ref{eq:totalshearstress}. As we can see, we do not obtain reasonable convergence until 50ETT, which is much longer than expected. It is clear that during the first 20 ETT, the simulation has a large statistical noise. 

In the case of the pipe flow, the differences among the four simulations appear to be small when analyzing the streamwise and wall-normal fluctuations, as displayed in Figure 3~(top), and the streamwise velocity, Figure 3~(bottom). As shown in the latter, the difference between the coarser and finer cases is below $0.3\%$ everywhere. Such accuracy, however, is insufficient for utilizing the indicator function $\Xi$ to extract reliable coefficients of the overlap region. As shown in Figure 3~(bottom), the different curves of $\Xi$ do not collapse exactly in the region between $y/\delta=0.3$ and $y/\delta=0.5$, where the slope value of $\Xi$ is critical. This is more clearly appreciated in Figure 5~(Top), where the differences across cases can be more clearly assessed. Notice that the finer mesh, PHRF case, corresponds to the solid red line. The other three cases have not converged to the PHRF's value even after 100 ETT. A better agreement is found in the case of the channels. Here, the convergence at $y/h=0.35$ is obtained after approximately 15 ETT. However, at $y/h=0.45$ the differences are of the order of 2\% after 80 ETT, which is an extremely lengthy computation for high-Reynolds-number simulations. 

The effect of this lack of adequate resolution can be better evaluated by the time-averaged values of the overlap coefficients $\kappa$, $B$, and $S_0$ as functions of ETT. After 50 ETTs, the differences among the cases are still above 3$\%$, as shown in Figure 6. Again, the convergence for the channel case is better, but there remains a small difference up to ETT of 150, which corresponds to very expensive computations. In any case, at least 20 ETTs are needed to obtain a minimum level of convergence. Such differences in the indicator function $\Xi$ should be smaller for any reliable simulation to predict the values of $\kappa$, $B$, and $S_0$. 

Finally, in the case of pipe flow, the values of these coefficients appear to depend on the azimuthal resolution, raising questions about the mesh aspect ratio of the grids. A careful study of such mesh variation effects in flows near their singularities, as in the case around the axis of pipes, is highly recommended. Such a study should also examine the sensitivity of the indicator function $\Xi$ to the order of the finite difference scheme used in the DNS. In general, the results demonstrate that classical and commonly used mesh sizes in the wall-normal direction for DNSs are not sufficiently fine to extract coefficients of the overlap region in wall-bounded turbulent flows, such as the von K\'arm\'an coefficient, $\kappa$. 

%


\begin{figure}
    \centering
    \begin{subfigure}{0.49\textwidth} 
    \includegraphics[width=0.99\textwidth]{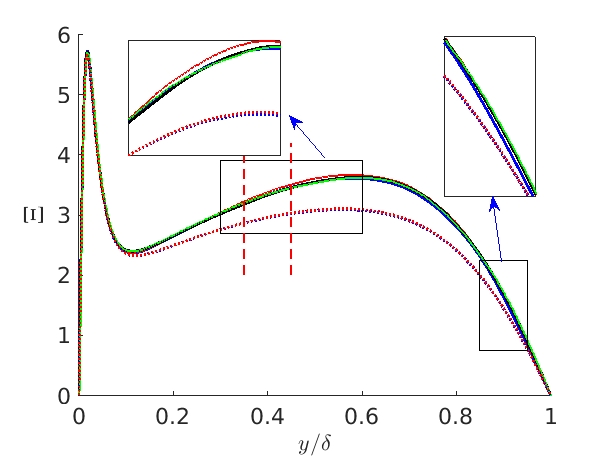}
    \end{subfigure}
    \begin{subfigure}{0.49\textwidth} 
    \includegraphics[width=0.99\textwidth]{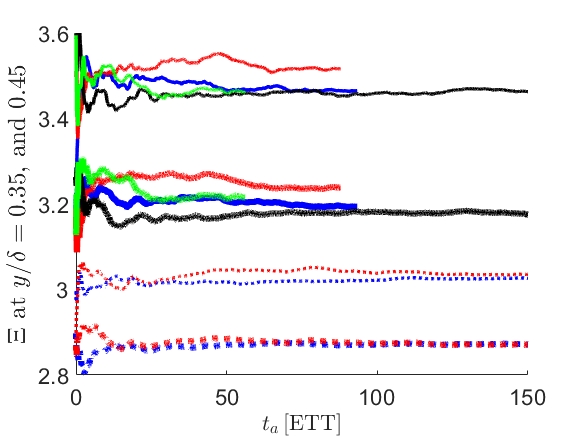}
    \end{subfigure}
    \label{fig:fig3}
        \caption{(Top) $\Xi$ for the six case under study. (Bottom) Values of $\Xi$ at $y/\delta=0.35$ and $0.45$ as a function of the averaging time $t_a$ expressed in eddy-turnover times (ETT).  Colors as in Table 1. Continuous lines for pipes, and dotted lines for channels. Thick lines represent, $y/\delta=0.35$, and thin ones, $y/\delta=0.45$.}
    \end{figure}
\begin{figure}
 \centering
        \begin{subfigure}{0.49\textwidth} 
    \includegraphics[width=0.99\textwidth]{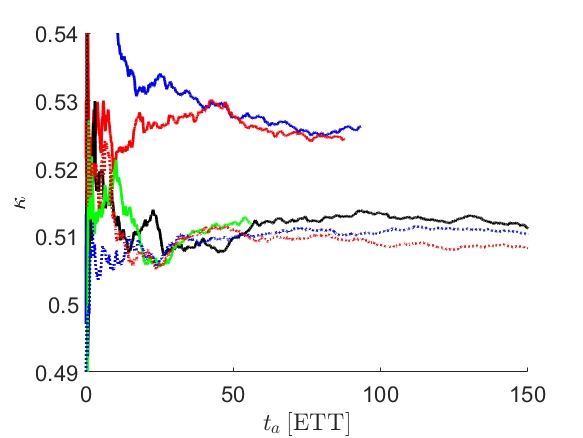}
    \end{subfigure}
    \begin{subfigure}{0.49\textwidth} 
    \includegraphics[width=0.99\textwidth]{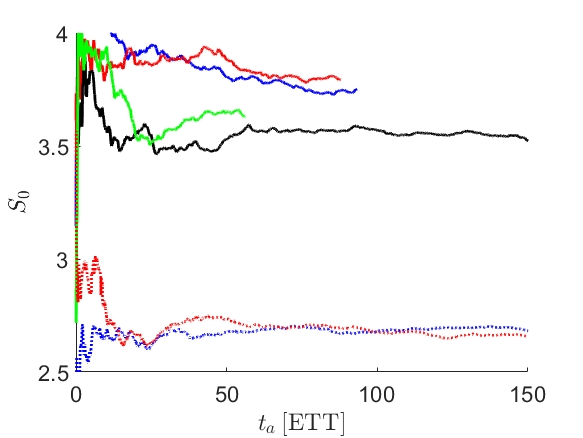}
    \end{subfigure}
    \caption{ $\kappa$ (top) and $S_0$ (bottom) for the six cases under study as functions of the averaging time $t_a$ expressed in ETT. Colors as in Table 1. Continuous lines for pipes, and dotted lines for channel}
    \label{fig:fig4}

\end{figure}


In conclusion, DNSs still stand as a powerful tool for exploring turbulent flows. However, the accuracy of DNS to compute certain important derivatives, such as those required for the indicator function $\Xi$ is challenged. To calculate $\Xi$, accurate values of the wall-normal derivative of the streamwise velocity must be established. This is a highly critical issue for experiments and should be addressed in all future measurements with more densely and equally spaced measurement locations in the wall-normal direction. Our investigation reveals that the indicator function exhibits sensitivity around the classic mesh size commonly used in DNSs. In particular, several coefficients of the overlap region required for Monkewitz and Nagib’s model of wall-bounded flows under streamwise pressure gradients cannot be found with less than 3\% difference among the various meshes we tested. For example, this creates a new uncertainty source in estimating the von K\'arm\'an coefficient, which is very important for modeling and predicting wall-bounded turbulent flows.  This uncertainty can also be generalized to any quantity needing wall-normal derivatives.\\

\textbf{Data availability:} The data used for this paper can be obtained by contacting S. Hoyas at serhocal@mot.upv.es. 

\section*{Acknowledgments}
For computer time, this research used the resources of the Supercomputing Laboratory at King Abdullah University of Science \& Technology (KAUST) in Thuwal, Saudi Arabia, Project K1652. RV acknowledges the financial support from ERC grant no. `2021-CoG-101043998, DEEPCONTROL'. Views and opinions expressed are however those of the author(s) only and do not necessarily reflect those of European Union or European Research Council. Neither the European Union nor granting authority can be held responsible for them.  SH is  funded by project PID2021-128676OB-I00 by Ministerio de Ciencia, innovación y Universidades / FEDER. 


\bibliographystyle{tsfp}
\bibliography{tsfp_six_pages}

\end{document}